\documentclass[superscriptaddress,preprintnumbers,reprint,prl,footinbib]{revtex4-1}
\usepackage{graphicx,amssymb,amsmath,color}
\usepackage{hyperref}

\newcommand{\beq}{\begin{equation}}
\newcommand{\eeq}{\end{equation}}
\def\bea{\begin{eqnarray}}
\def\eea{\end{eqnarray}}
\newcommand{\bei}{\begin{itemize}}
\newcommand{\eei}{\end{itemize}}

\newcommand{\Fig}[1]{Fig.~\ref{#1}}
\newcommand{\Eq}[1]{Eq.~(\ref{#1})}

\def\={\,=\,}
\def\+{\,+\,}
\def\-{\,-\,}
\def\Mch{{\cal M}}
\def\Msun{M_\odot}

\begin{document}

\title{Gravitational-Wave Fringes at LIGO: Detecting Compact Dark Matter by Gravitational Lensing}

\author{Sunghoon Jung}
\email{sunghoonj@snu.ac.kr}
\affiliation{Center for Theoretical Physics, Dept. of Physics and Astronomy, Seoul National University, Seoul 08826, Korea}

\author{Chang Sub Shin}
\email{csshin@ibs.re.kr}
\affiliation{Center for Theoretical Physics of the Universe, IBS, Daejeon 34051, Korea}

\begin{abstract}
Utilizing gravitational-wave (GW) lensing opens a new way to understand the small-scale structure of the universe.
We show that, in spite of its coarse angular resolution and short duration of observation, LIGO can detect the GW lensing induced by compact structures, in particular by compact dark matter (DM) or primordial black holes of $10 - 10^5 \, \Msun$, which remain interesting DM candidates. The lensing is detected through GW frequency chirping, creating the natural and rapid change of lensing patterns: \emph{frequency-dependent amplification and modulation} of GW waveforms. As a highest-frequency GW detector, LIGO is a unique GW lab to probe such light compact DM. 
With the design sensitivity of Advanced LIGO, one-year observation by three detectors can optimistically constrain the compact DM density fraction $f_{\rm DM}$ to the level of a few percent.
\end{abstract}

\preprint{CTPU-17-41}

\maketitle


{\bf Introduction.} 
The GW from far-away binary mergers~\cite{Abbott:2016blz,TheLIGOScientific:2017qsa} is a new way to see the universe with gravitational interaction. Not only is it revealing astrophysics of solar-mass black holes and neutron stars, but the GW can also carry information of intervening masses and the evolution of the universe through gravitational lensing. Having the long wavelength $\lambda$, the GW is usually expected to be lensed by heaviest structures (such as galaxies and their clusters) with large enough Schwarzschild radii, $2 G M /c^2 = 2M \gg \lambda \simeq 2 \times 10^3 \, (100 \,{\rm Hz}/f) \, \Msun$. Their prototypical lensing signal is strongly time-delayed GW images~\cite{Sereno:2010dr,Piorkowska:2013eww} or statistical correlations~\cite{Laguna:2009re}.

In this letter, we show that LIGO can detect the GW lensing induced by much lighter compact  structures. The new lensing observable depends crucially on the GW frequency evolution during binary inspiral -- ``chirping''. The chirping provides the natural and rapid change of lensing patterns so that LIGO can detect relatively weak GW lensing, in spite of its coarse angular resolution (${\cal O}(1-10)$ deg~\cite{Aasi:2013wya,Graham:2017lmg}, let alone typical strong-lensing image separations of arcsec) and short measurement time (less than minutes, let alone typical micro-lensing observations of a few weeks). Remarkably, measuring highest frequencies of $f = 10 - 5000$ Hz, LIGO is a unique GW detector to see compact structures as light as $M = 10 - 10^5 \Msun$.

An important example of such light structure is the compact DM. It remains an attractive DM candidate, predicted by various models of particle physics and cosmology: axion miniclusters, compact mini halos, and primordial black holes~\cite{Hawking:1971ei,Carr:1974nx,Kolb:1993zz,Kolb:1995bu,Bringmann:2001yp,Blais:2002gw,Berezinsky:2003vn,Diemand:2005vz,Zurek:2006sy,Ricotti:2009bs,Kopp:2010sh,Hardy:2016mns}. Compact DM is mainly searched by light lensing: micro-lensing (temporal variation of brightness)~\cite{Tisserand:2006zx,Novati:2013fxa} and strong-lensing (multiple images)~\cite{Nemiroff:1990,Wilkinson:2001vv}. But, in a wide range of compact DM mass $10^{-16} - 10^5 \Msun$, its density fraction $f_{\rm DM} \lesssim 0.1 -1$ of total DM density remains to be probed~\cite{Carr:2016drx}. We present the prospect for the new LIGO lensing observable to probe the compact DM of $M_{\rm DM} = 10 -10^5 \Msun$.

\begin{figure}[t] \centering
\includegraphics[width=0.46\textwidth]{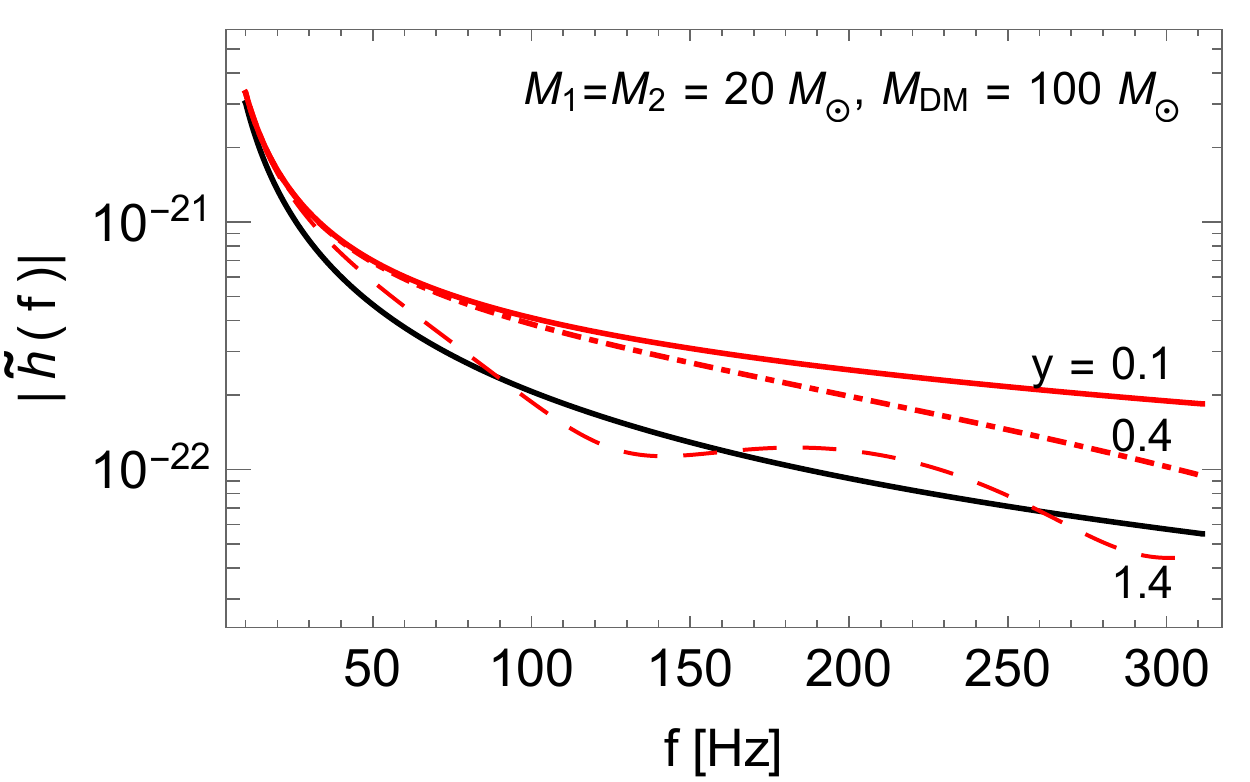}
\caption{Illustration of GW lensing fringe at aLIGO: Lensed (red) vs. unlensed (black)  waveforms. The lensing compact DM mass $M_{\rm DM} = 100 \, \Msun$ and redshifted GW binary masses $M_1 = M_2 = 20 \,\Msun$ merging at $f \simeq 320$ Hz. At  small impact angle $y = \theta_s / \theta_E$ (red-solid) \Eq{eq:lensed}, \emph{frequency-dependent amplification} cannot be fit by a constant rescaling $A_0$ (\Eq{eq:A0}) of unlensed waveforms (black); whereas, at large $y$ (dashed), \emph{frequency-dependent modulation} cannot be matched by a constant phase-shift $\phi_0$ (\Eq{eq:phi0}). 
In general (dotdashed), both lensing effects co-exist.}
\label{fig:lensed}
\end{figure}


{\bf GW Lensing Observable.} 
The proposed GW lensing observable relies on the following properties (in contrast to those of light).

Above all, the binary GW frequency chirps. 
Suppose that some lensing creates two GW images with (small) time-delay $\Delta t_d$. The two images interfere since LIGO cannot resolve them. Then, the phase-shift between them $\sim f \Delta t_d$ grows with the frequency, and the resulting interference pattern \emph{changes} with chirping~\cite{Nakamura:1997sw,Takahashi:2016jom} (see \Fig{fig:lensed} dashed) -- frequency-dependent modulation. The final stage of binary inspiral (observed by LIGO) is where the frequency-dependent signal can be detected most efficiently because the frequency is highest and grows most rapidly. 

Secondly, the GW wavelength is much longer than that of light. 
Therefore, GW lensing does not always produce two images with constant time-delay. In general, the lensed GW waveform, $\widetilde{h}^L(f)$, is a superposition of all unlensed waveforms, $\widetilde{h}(f)$, that follow all possible null rays (passing $\vec{\theta}$ in the thin lens plane at redshift $z_{\rm DM}$)~\cite{Schneider:1992,Takahashi:2003ix}
\beq
\widetilde{h}(f)^L = \tfrac{d_L d_S (1+z_{\rm DM})}{i d_{LS}}f \int d^2 \vec{\theta}   \exp\big[i2\pi f \Delta t( \vec{\theta},\vec{\theta}_s) \big]  \, \widetilde{h}(f),
\label{eq:lensed} \eeq
with $d_{L,S,LS}$ the angular-diameter distance to the lens, source and between them; the compact DM is treated as a point lens~\cite{Schneider:1992,Takahashi:2003ix} with the time-delay $\Delta t(\vec{\theta},\vec{\theta}_s)$ 
$= (1+z_{\rm DM}) \left( \frac{d_L d_S}{2 d_{LS}} | \vec{\theta}-\vec{\theta}_s|^2- 4M_{\rm DM}|\vec{\theta}| \right)$
and the source impact angle $y= \theta_s / \theta_E$ normalized by the Einstein angle $\theta_E = \sqrt{ \frac{4 GM_{\rm DM}}{c^2} \frac{d_{LS}}{d_L d_S} }$. 
Only when the GW frequency is larger than the inverse of the typical time-delay between null rays $f \Delta t_d \simeq 4f M_{\rm DM} \simeq 2 \times 10^{-5} (M_{\rm DM}/\Msun) (f/ {\rm Hz}) \gg 1$ (equivalently, $\lambda \ll M_{\rm DM}$), the integral is dominated by discrete stationary points with separate images (geometric optics limit). In the opposite limit, GW diffraction becomes important and lensing amplification becomes weaker (wave optics limit); eventually, the GW does not see the lens if its wavelength becomes too long.

In particular, the GW wavelength in the LIGO band, $\lambda \simeq 2 \times 10^3 \, (100 \,{\rm Hz}/f) \, \Msun$, is comparable to the Schwarzschild radius of the compact DM with $M_{\rm DM}=10-10^5 \Msun$. Chirping from 10 Hz to ${\cal O}(100-1000)$ Hz, GW lensing by such masses may transition from wave-optics ($\lambda_{\rm GW} \gtrsim {\cal O}(M_{\rm DM})$) to geometric-optics ($\lambda_{\rm GW} \lesssim {\cal O}(M_{\rm DM})$). Therefore, with chirping, GW lensing magnitude (both amplification and time-delay) also grows; compare low and high frequency regions in \Fig{fig:lensed}. 

The two effects combined -- \emph{frequency-dependent amplification and modulation} -- define our ``GW fringe'' lensing signal. \Fig{fig:lensed} illustrates fringes in comparison to the unlensed waveform. Below, we will calculate the fringe detection efficiency at LIGO, lensing optical depth, detection rate, and expected constraints on $f_{\rm DM}$.



{\bf Analysis for Lensing Detection.} 
In the LIGO frequency band,  binary GWs spend only a few seconds or minutes. Therefore, a detector on the Earth is almost at rest during the measurement. Then, the unlensed GW waveform is sinusoidal when observed by a single LIGO detector
\bea
h(t) &=& A_+(t)F_+ \cos \phi(t)  \+ A_\times (t)F_\times \sin \phi(t) \\
&=& A(t) \cos \,( \, \phi(t) \+ \phi_0 \,),  \nonumber
\eea
with detector response functions $F_{+,\times}$ constant in time during the measurement period. Rather, the time dependences of amplitude $A(t)$ and phase $\phi(t)$ are determined uniquely by the redshifted chirp mass $\Mch_z = (M_1^3 M_2^3/(M_1+M_2))^{1/5}$ (with redshifted $M_1=M_2$ in this work):
$A(t) \propto {\cal M}_z^{5/3} f(t)^{2/3}$ with $f'(t) = \frac{96}{5} \pi^{8/3} f(t)^{11/3} \Mch_z^{5/3}$ at the leading Newtonian order, where $f(t) \equiv \phi^\prime(t) / 2\pi$ is the instantaneous GW frequency.

Now, the unlensed waveform observed in the frequency domain is simply
\beq
\widetilde{h}(f) \= \widetilde{A}(f) \exp[ i (\Psi(f) + \phi_0) ],
\label{eq:phi0} \eeq
where
\bea
\widetilde{A}(f) &=& A_0 \, \Mch_z^{5/6} f^{-7/6},
\label{eq:A0} \\
\Psi(f) &=& 2\pi f t_c + \frac{3}{128} ( \pi {\cal M}_z f )^{-5/3},
\eea
and any constant phase-shift is absorbed into $\phi_0$. Although the real-valued constant $A_0$ depends on various source parameters (such as polarization and distance), single short measurement can only measure $A_0$.
For simplicity, we ignore higher-order post-Newtonian terms, spin effects, orbital-eccentricity, and non-quadrupole modes; see~\cite{footnote1} for caveats.


\begin{figure}[t] \centering
\includegraphics[width=0.45\textwidth]{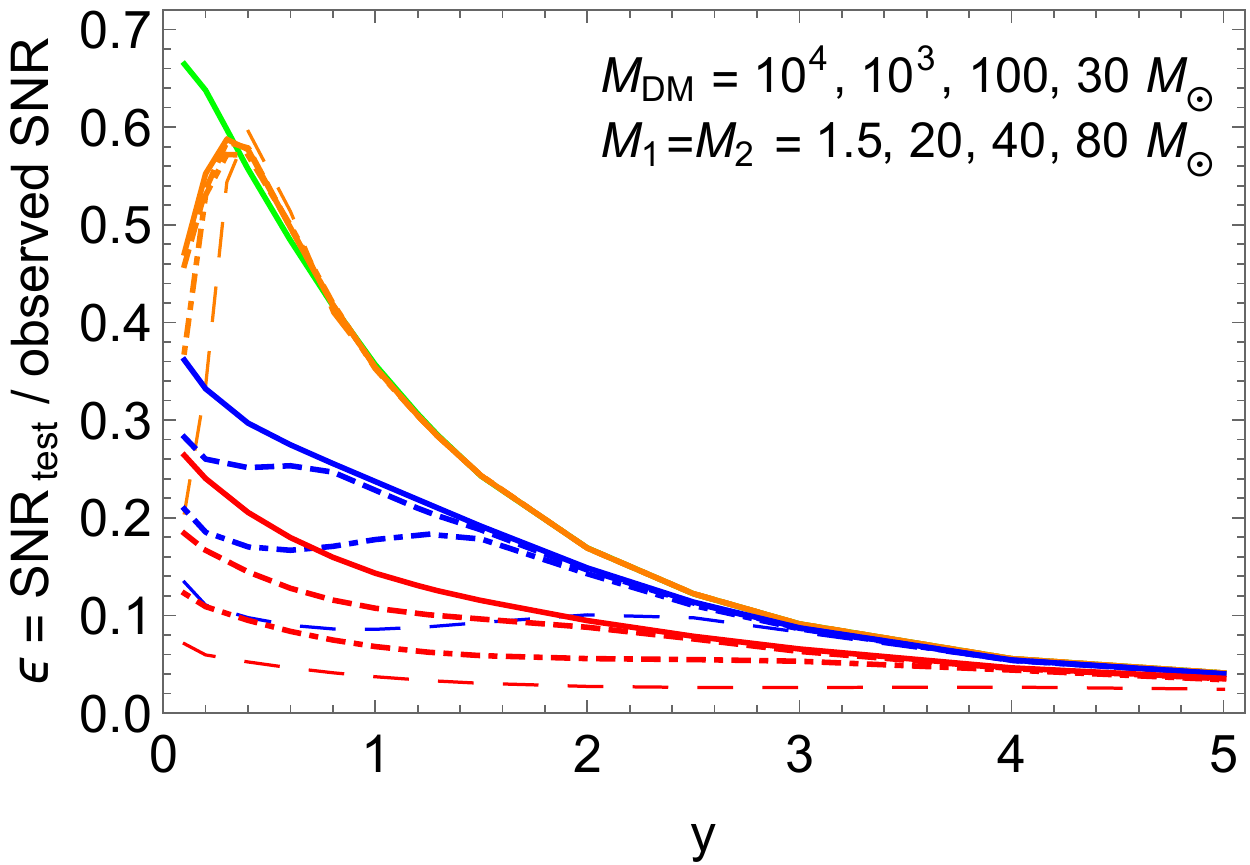}
\caption{The lensing-detection efficiency $\epsilon$ in \Eq{eq:eps} of a single aLIGO detector with design sensitivity. The DM masses are $M_{\rm DM} = 10^4 $ (green), $10^3$ (orange), 100 (blue), 30 (red) $\Msun$ and redshifted binary masses $M_1=M_2 =$ 1.5 (solid), 20 (dashed), 40 (dot-dashed), 80 (long-dashed) $\Msun$.}
\label{fig:testSNRratio}
\end{figure}

We use the $\chi^2$ least-squares fit method to determine whether GW lensing can be detected~\cite{footnote2}.
We define the goodness of fit of (trial) unlensed waveforms to the (observed) lensed waveform as
\beq
({\rm SNR}_{\rm \, test})^2 \, \equiv\,  4 \int_{f_0}^{f_1} \, \frac{ \big| \widetilde{h}(f)^L - \widetilde{h}(f)_{\rm best-fit} \big|^2 }{ S_n (f) } \, df, 
\label{eq:chi} \eeq
similarly to the observed signal-to-noise ratio (SNR)
\beq
({\rm SNR})^2 \= 4 \int_{f_0}^{f_1} \, \frac{ | \widetilde{h}(f)^L|^2 }{ S_n(f) } df.
\label{eq:snr} \eeq
The integration is done over the aLIGO frequency band~\cite{TheLIGOScientific:2016agk}:
$f_0 =$ 10 Hz, $f_1 =$ min($f_{\rm cut}, 5000 \, {\rm Hz})$,
where the cut-off frequency $f_{\rm cut}=(3\sqrt{3} \pi (M_1+M_2))^{-1}$ at $r=3 (M_1 + M_2)$~\cite{footnote3}. 

For the given observed lensed-waveform $\widetilde{h}(f)^L$, we find the best-fit unlensed waveform $\widetilde{h}(f)_{\rm best-fit}$ by minimizing ${\rm SNR}_{\rm \, test}$ over two fitting parameters, $A_0$ and $\phi_0$. Here, we assume that the chirp mass $\Mch_z$ and the coalescence time $t_c$ (time at which $f$ formally diverges) are well measured regardless of lensing effects~\cite{Cutler:1994ys} (but see also, e.g.,~\cite{Cao:2014,Dai:2016igl}).  
The simplified best-fit based on $A_0$ and $\phi_0$ is convenient to capture the leading physics of GW fringe, as illustrated in \Fig{fig:lensed}; but see~\cite{footnote1} for caveats.

We regard that the existence of GW lensing is detected if ${\rm SNR}_{\rm \, test} \geq 3$ or 5 with ${\rm SNR} \geq 8$~\cite{footnote4}. 
With multiple detectors, we require the total SNR quadrature-sum to satisfy these criteria.
In \Fig{fig:testSNRratio}, we show the lensing-detection efficiency of single aLIGO detector defined as
\beq
\epsilon \, \equiv \, \frac{ {\rm SNR}_{\rm \, test} }{ {\rm SNR} }.
\label{eq:eps} \eeq
It is typically $\sim {\cal O}(10)\%$ so that strong GWs with SNR $\sim {\cal O}(10)$ can be lensing-detected with ${\rm SNR}_{\rm \, test} \sim {\cal O}(1)$. The heavier the lens, the larger $\epsilon$, trivially. The lighter GW binaries, the larger $\epsilon$ because lighter GWs merge at higher frequencies experiencing more modulation.

The slight fluctuation and drop of the $\epsilon$ curve in \Fig{fig:testSNRratio} are related to the interplay of frequency-dependent modulation and amplification.
The former becomes more significant at large-$y$ due to larger time-delay, whereas the latter at small-$y$ due to stronger lensing; see \Fig{fig:lensed}.
Intermediate regimes may produce less exotic waveforms (e.g., dot-dashed in \Fig{fig:lensed}), if the GW merges just before the first destructive interference.


{\bf Advanced LIGO Prospects.} 
We turn to discuss expected results from three aLIGO detectors, representing a network of future detectors. We assume design aLIGO sensitivity (the dark-blue noise curve in Fig.1 of Ref.~\cite{TheLIGOScientific:2016agk} and horizon in Ref.~\cite{Martynov:2016fzi}), yielding $\sim3$ times larger SNR (seeing $\sim3$ times farther distance) than current LIGO.

\begin{figure}[t] \centering
\includegraphics[width=0.45\textwidth]{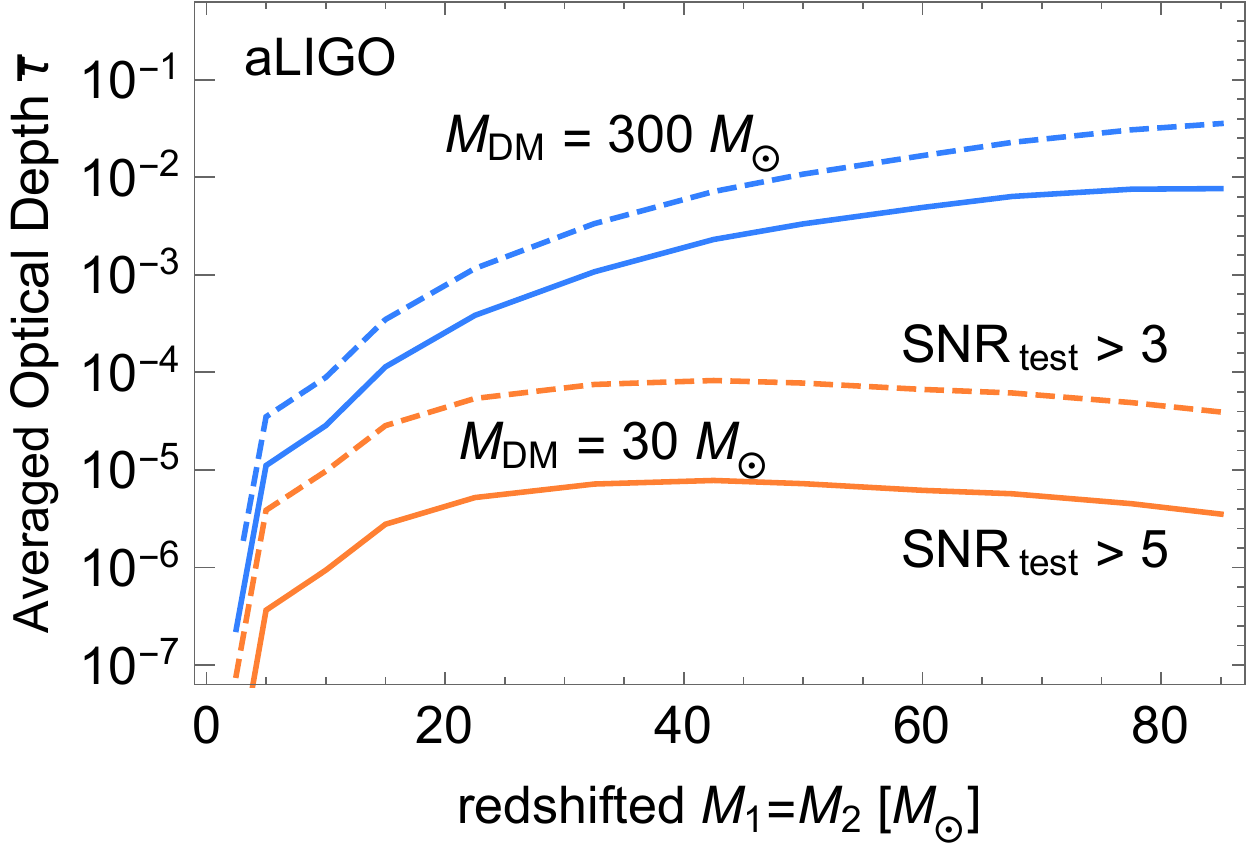}
\caption{The volume-averaged optical depth, $\bar{\tau}$, as defined in \Eq{eq:intoptdepth}. It is the probability for detectable lensing, including the probability for ${\rm SNR}_{\rm test} > $3 (dashed) or 5 (solid) and ${\rm SNR}>8$ from three aLIGOs. The upper (lower) set of curves is for $M_{\rm DM} = 300$ (30) $\Msun$. $f_{\rm DM}=1$.}
\label{fig:int_optdepth}
\end{figure}


The differential optical depth for \emph{detectable} lensing for the given lensing parameters (source and lens masses and locations -- ${\cal M}_z, z_S$ and $M_{\rm DM}, z_{\rm DM}, y$) is
\bea
\frac{d^2\tau}{dy \, dz_{\rm DM}} 
&=&  \frac{c a(z_{\rm DM})}{H(z_{\rm DM})} \, 2 \pi (d_L \theta_E)^2 y \, (1+z_{\rm DM})^3 n_{\rm DM} \,P(w) \label{eq:diffoptdepth} \\
&=& \frac{3}{2} f_{DM} \Omega_{DM} \,  \frac{H_0^2 (1+ z_{\rm DM})^2 }{H(z_{\rm DM}) c} \frac{d_L d_{LS} }{d_S} \, 2y \,P(w), \nonumber
 \eea 
where a constant comoving lens density $n_{\rm DM}$ is assumed for the compact DM mass density $f_{\rm DM} \Omega_{\rm DM}$.
 The optical depth depends on the detectability. 
The parameter-dependent detection efficiency $\epsilon$ (\Eq{eq:eps}) determines the minimum SNR needed for detectable lensing: max(8, $3/\epsilon$ or $5/\epsilon$) depending on the SNR$_{\rm test}>3$ or 5 criteria.
Then, among the sources with randomly chosen sky position, inclination and polarization angle, the probability for SNR to be greater than the minimum value is denoted by $P(w)$ with $w =$ (SNR needed)/(SNR optimal), where (SNR optimal) is the maximum possible SNR for some optimally oriented source. $P(w)$ is the cumulative distribution of $w$ spanned by such random source parameters; $P(w)=1$ for $w=0$, and decreases to $P(w)=0$ for $w>1.4$ for three detectors~\cite{Dominik:2014yma}.
The probability is convoluted with comoving volume and lens density to yield the optical depth $\tau$ (for the given ${\cal M}_z, z_S$ and $M_{\rm DM}$). We assume a standard cosmology with the Hubble parameter $H(z)$, mass density fractions $\Omega_m = 0.27, \, \Omega_{DM} = 0.24$, $\Omega_\Lambda = 0.73$ for matter, DM and cosmological constant.

The source distribution -- the comoving merger-rate density $\dot{n}_{\rm merger}$ -- is assumed to be constant in $z_S$ for simplicity~\cite{footnote5}, but its dependence on ${\cal M}_z$ is kept. Two sets of distributions on ${\cal M}_z$ are taken from Ref.~\cite{Belczynski:2016obo}: the optimistic merger model M1 predicting highest merger rate consistent with LIGO's observations, and the pessimistic model M3. 
For the given masses ${\cal M}_z$ and $M_{\rm DM}$, the $\dot{n}_{\rm merger}({\cal M}_z)$ yields the merger rate $\dot{N}_{\rm merger} ({\cal M}_z) = \int dV_{\rm com} \, \dot{n}_{\rm merger}({\cal M}_z)$ and, finally, the detectable lensing rate as 
\beq
\dot{N}_{\rm lensing}({\cal M}_z, M_{\rm DM}) = \int dV_{\rm com} \, \dot{n}_{\rm merger} \, \tau(z_S),
\label{eq:lensingrate}\eeq
where $\tau(z_S) =\int_0^{z_S} dz_{\rm DM} \int dy \frac{d\tau}{dy dz_{\rm DM}}$ and the comoving volume element $dV_{\rm com} = \frac{4\pi \chi(z_S)^2c}{H(z_S)} dz_S$ in terms of the comoving distance $\chi$.

In \Fig{fig:int_optdepth}, we show the ratio of detectable lensing to the total merger as the volume-averaged optical depth (again for the given masses ${\cal M}_z$ and $M_{\rm DM}$)
\beq
\bar{\tau}({\cal M}_z, M_{\rm DM})  \equiv  \frac{\dot{N}_{\rm lensing}}{\dot{N}_{\rm merger}} = \frac{ \, \int d z_S \, \frac{4\pi \chi(z_S)^2 c}{H(z_S)} \, \tau(z_S) }{ \int d z_S \, \frac{4\pi\chi(z_S)^2 c}{H(z_S)} }.
\label{eq:intoptdepth} \eeq
The averaged optical depth $\bar{\tau}$ is a function (only) of ${\cal M}_z$ and $M_{\rm DM}$. 
It is smallest for lightest binaries mainly because GWs are weakest. 
It becomes largest at ${\cal M}_z \sim 30 \Msun$ for light $M_{\rm DM}$ as $\epsilon$ is smaller for heavier binaries, whereas it keeps growing for heavy $M_{\rm DM}$ since $\epsilon$ depends less on the binary mass; see \Fig{fig:testSNRratio}.

\begin{figure}[t] \centering
\includegraphics[width=0.45\textwidth]{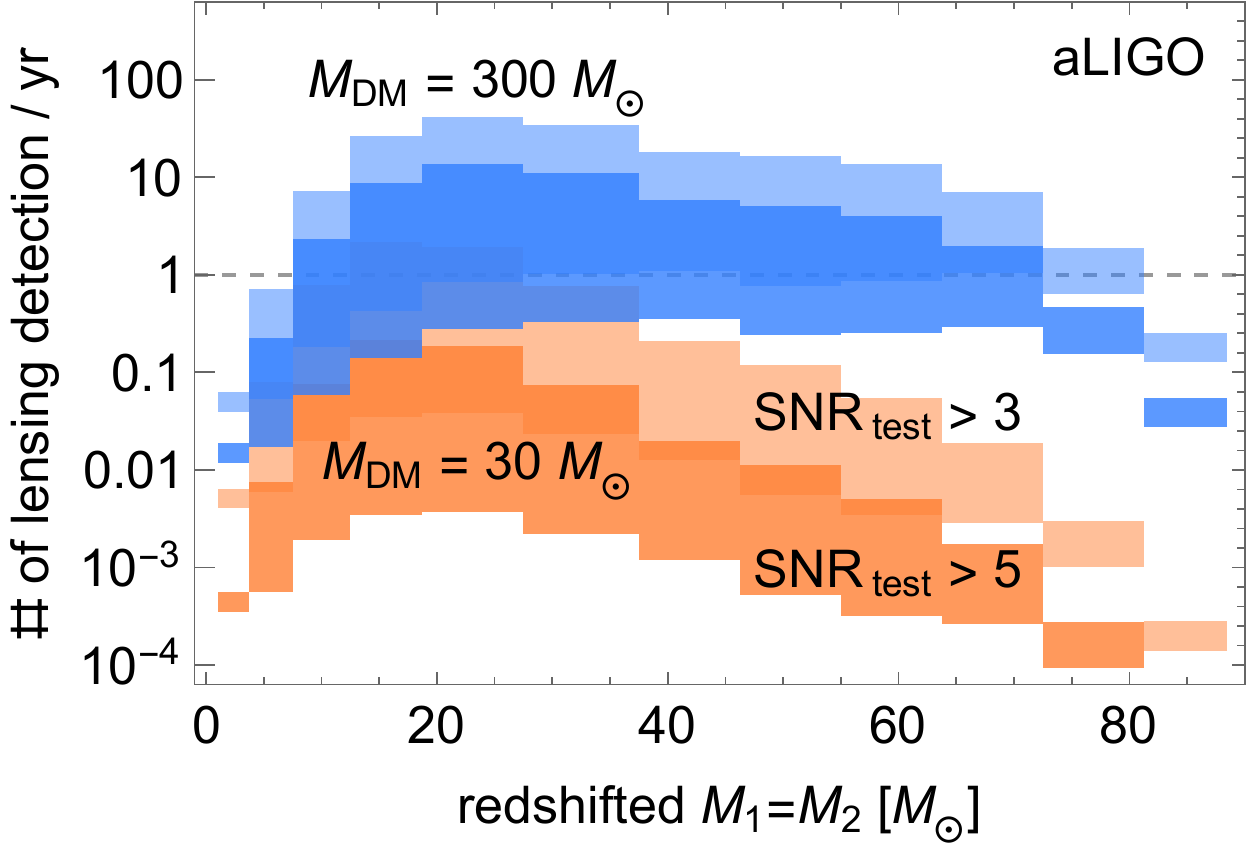}
\caption{The number of GW lensing detections per year by three aLIGOs. The darker (lighter) region satisfies ${\rm SNR}_{\rm \, test}>$ 5 (3) and ${\rm SNR} > 8$, and vertical ranges span between the optimistic merger-rate M1 and the pessimistic M3~\cite{Belczynski:2016obo}. The blue (orange) region is for the lensing $M_{\rm DM} = 300$ (30) $\Msun$. The horizontal dashed line denotes one detection per year. The total detection rate is the sum over mass bins. $f_{\rm DM}=1$.}
\label{fig:lensingdetection-Mch}
\end{figure}

Shown in \Fig{fig:lensingdetection-Mch} is the final lensing detection rate, $\dot{N}_{\rm lensing}$ in \Eq{eq:lensingrate}, from three aLIGO detectors. The highest detection rate is expected from ${\cal M}_z = 20-30 \Msun$, as $\bar{\tau}$ is typically largest there. Total expected yearly detections are sizable: optimistically, 6 (170) for $M_{\rm DM} = 30 \,(300) \Msun$ with ${\rm SNR}_{\rm test} >3$ and 0.6 (55) with ${\rm SNR}_{\rm test} >5$. Pessimistic expectations are about 25-40 times smaller.

\begin{figure}[t] \centering
\includegraphics[width=0.45\textwidth]{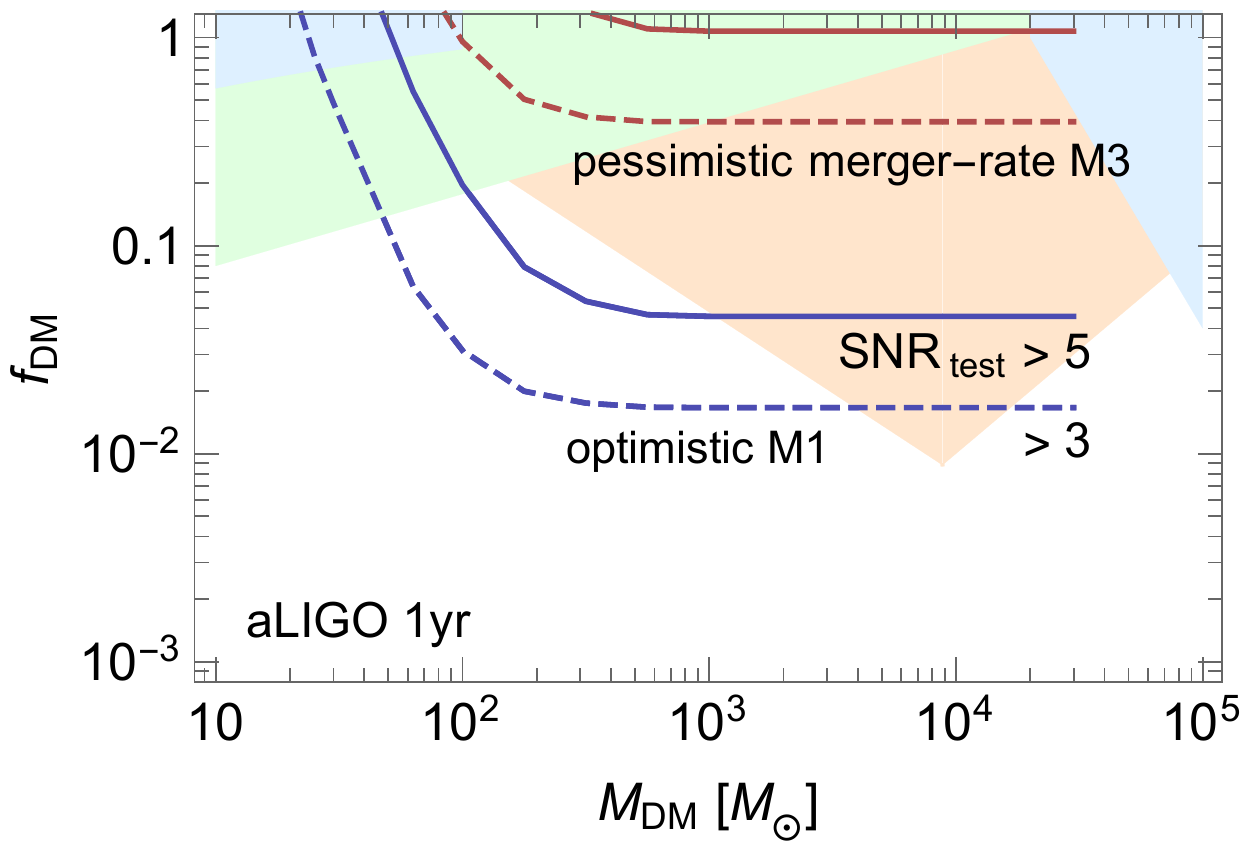}
\caption{Expected 95\% CL constraints on $f_{\rm DM}$ from one-year observation by three aLIGOs. The Poisson distribution of fringe detection probability is assumed.
 ${\rm SNR}_{\rm \, test} > 3$ (dashed) or 5 (solid) is used with optimistic (blue) or pessimistic (red) merger-rate models. Shaded regions are existing constraints: microlensing (left blue)~\cite{Novati:2013fxa}, caustic crossing (green)~\cite{Oguri:2017ock}, cluster survival (orange)~\cite{Brandt:2016aco}, and millilensing of quasars (right blue)~\cite{Wilkinson:2001vv}.}
\label{fig:fDM}
\end{figure}

Based on the Poisson distribution of the number of fringe detections, we calculate the value of $f_{\rm DM}$ giving 95\% probability of one detection, shown in \Fig{fig:fDM} as 95\% CL constraints on $f_{\rm DM}$ assuming null detections. But a proper characterization of the probabilitiy distribution will be needed to derive more realistic constraints.
The sensitivities start from $M_{\rm DM} \sim 10 \Msun$, become strongest for $M_{\rm DM} \gtrsim 10^2 \Msun$, and stop shown for $M_{\rm DM} \gtrsim 10^5 \Msun$. The three regions have different values of the phase-shift
\beq
f \Delta t_d \simeq 2\times 10^{-5} (M_{\rm DM}/\Msun) (f/{\rm Hz}) \label{eq:fdeltat}
\eeq
(in the wave-optics regime, we can think of $\Delta t_d$ as a typical time-delay from null rays with $\theta \simeq \theta_E$).

For $f \Delta t_d  \gtrsim 0.1$ from $M_{\rm DM} \gtrsim 10^2 \,\Msun$, the GW fringe is most pronounced, and LIGO fringe search is a powerful probe of those compact DM. Here, the ${\cal O}(1)$ evolution of the frequency in the LIGO band produces $\gtrsim {\cal O}(1)$ cycles of fringe patterns, which is easiest to detect. Resulting optimistic sensitivity $f_{\rm DM} \lesssim 10^{-2}$ is comparable to or stronger than existing constraints from mircolensing~\cite{Novati:2013fxa}, millilensing~\cite{Wilkinson:2001vv}, star's caustic-crossing~\cite{Oguri:2017ock,Diego:2017drh,Venumadhav:2017pps}, and star-cluster survival~\cite{Brandt:2016aco} as well as various proposed searches~\cite{Munoz:2016tmg,Zumalacarregui:2017qqd}. 
The GW fringe sensitivity will further improve with longer observation time.

The strong sensitivity is attributed to high detection efficiencies $\epsilon$ and large merger rates. Not only can LIGO see sources at far distance ($z_S \lesssim 2$ in this paper, but farther with future upgrades), but also at large $y$ too. Remarkably, large $y \gtrsim 1.5$ can still lead to sizable efficiency $\epsilon = 5-25\%$ in \Fig{fig:testSNRratio}. Although lensing is weak there, the GW \emph{amplitude} modification can still be $\sim$10\% (4\%) for $y=$ 3 (5). Combined with frequency-dependent interference over a range of frequencies, this can lead to such sizable $\epsilon$. On the contrary, light lensing is observed through its brightness (squared amplitude) features so that large-$y$ lensing effects are hardly observable.

The constraints become almost constant for heavy masses $M_{\rm DM} \gtrsim 200 \Msun$ in \Fig{fig:fDM} because the decrease of the DM number density (with heavier DM) is compensated by the increase of the Einstein radius in the last line of \Eq{eq:diffoptdepth}. 
Nevertheless, we stop showing the result at $M_{\rm DM} \sim 10^{5} \Msun$, since waveform modulations with $f \Delta t_d \gtrsim 10^3$ in the heavy-$M_{\rm DM}$ region maybe too quick to be temporally resolved.
On the other hand, LIGO fringe search is not sensitive to $M_{\rm DM} \lesssim {\cal O}(10) \Msun$. Here, small phase-shift $f \Delta t_d \lesssim 0.1$ barely produces observable fringes. 

In general, a lensing fringe becomes most pronounced when the following relation is satisfied
\beq
(M_{\rm DM}/\Msun )(f/{\rm Hz}) \gtrsim 10^4 - 10^6,
\label{eq:fringecondition} \eeq
equivalent to $f \Delta t_d \gtrsim 0.1 - 10$ from \Eq{eq:fdeltat}. The maximum $f \Delta t_d$ (hence, $M_{\rm DM} f$) depends on the detector's temporal resolution, as discussed. But as a highest-frequency GW detector, LIGO can see lightest DM;
lower-frequency detectors (such as LISA) can probe only heavier DM.
This discussion also applies to the photon fringe. The compact DM of $10^{-16}-10^{-14} \Msun$ is expected to produce (femto-lensing) fringes on gamma-ray bursts (GRBs) with $f_{\rm GRB} \simeq 10-1000$ keV $\simeq 2.4 \times (10^{18} - 10^{20})$ Hz~\cite{Gould:1992}, satisfying \Eq{eq:fringecondition} and (\ref{eq:fdeltat}). Although the GRB frequency does not change with time, a fringe spectrum can be observed~\cite{Barnacka:2012bm}.


{\bf Conclusion.}
We have shown that LIGO can detect the GW lensing fringe induced by the compact DM of $M_{\rm DM} = 10-10^5 \Msun$. The LIGO measurement of GW fringes can surpass or strengthen existing constraints on such DM fraction, as shown in \Fig{fig:fDM}. Without LIGO fringe measurements, this small structure could not have been probed with GW. 
The general relation \Eq{eq:fringecondition} suggests that lower-frequency detectors can probe heavier compact DM. Therefore, future broadband GW surveys covering $f= 10^{-9} - 10^3$ Hz (from various detectors) are needed to probe a wide range of structures through the GW fringe.


\emph{Acknowledgements.} We thank Hyung Mok Lee and Hyung Won Lee for useful comments. SJ is supported by Korea NRF-2017R1D1A1B03030820 and NRF-2015R1A4A1042542. CSS is supported by IBS under the project code IBS-R018-D1.



\begin{thebibliography}{99}
\bibitem{footnote1}
Spin precession, orbital eccentricity and non-quadrupole modes can introduce amplitude oscillations. Higher-order post-Newtonian effects can introduce a frequency-dependent amplitude scaling $\propto f^{-7/6}(1+\Sigma_k \alpha_k f^{(k-5)/3})$; some values of $\alpha_k(m_1, m_2, s_1, s_2)$ might provide a good agreement with lensed waveform. A more dedicated study should include these effects.
\bibitem{footnote2}
For example, a relative likelihood
$\protect\langle \tilde{h}^L \protect| \tilde{h}^L \protect\rangle - \protect\langle \tilde{h}^L \protect| \tilde{h}_{ \protect\rm best-fit} \protect\rangle^2 / \protect\langle \tilde{h}_{\protect\rm best-fit} \protect| \tilde{h}_{\protect\rm best-fit } \protect\rangle$ more directly
measuring how much lensed waveform improves best-fit can be used, where $\protect\langle \tilde{h}^L \protect| \tilde{h}^L \protect\rangle$ is defined to be Eq.~\ref{eq:snr}. Once we find the $\tilde{h}_{\protect\rm best-fit}$ by minimizing the likelihood, we obtain very similar numerical results, e.g., Fig.~\ref{fig:testSNRratio}. Mathematically, the minimization actually makes both observables identical.
\bibitem{footnote4}
The probability for random noise to mimic lensing distortion (on top of unlensed waveform) would be $\sim \exp( - {\rm SNR}_{\rm \, test}^2/2 ) \lesssim 3 \times 10^{-6}, \, 0.01$ for ${\rm SNR}_{\rm \, test} \geq 5, \, 3$, respectively. The former probability is much smaller than typical strong lensing probability shown in \Fig{fig:int_optdepth} so that ${\rm SNR}_{\rm \, test} \geq 5$ can be confidently associated to the GW lensing, whereas ${\rm SNR}_{\rm \, test} \geq 3$ can be somewhat loose.
\end{thebibliography}
\end{document}